\documentclass[letterpaper]{article}
\usepackage{natbib,alifeconf}

\title{Divergent Cumulative Cultural Evolution}
\author{Chris Marriott$^{1}$ \and Jobran Chebib$^2$ \\
\mbox{}\\
$^1$University of Washington, Tacoma, WA, USA 98402 \\
$^2$University of Z\"urich, Z\"urich, Switzerland 8057 \\
dr.chris.marriott@gmail.com}

\begin{document}
\maketitle

\begin{abstract}
Divergent cumulative cultural evolution occurs when the cultural evolutionary trajectory diverges from the biological evolutionary trajectory.  We consider the conditions under which divergent cumulative cultural evolution can occur.  We hypothesize that two conditions are necessary.  First that genetic and cultural information are stored separately in the agent.  Second cultural information must be transferred horizontally between agents of different generations.  We implement a model with these properties and show evidence of divergent cultural evolution under both cooperative and competitive selection pressures. \end{abstract}
\section{Introduction}
Social learning is a form of learning that arises from social \textit{situatedness} \citep{LZ} and is characterized by agents interacting with one another in order to learn.  Social learning can accelerate learning beyond that of individual learning strategies (see \cite{MC14,MPD10}) and most notably is its ability to support cumulative cultural evolution \citep{W11,MWL,hm03,br96}.  Cumulative cultural evolution is an adaptive process in which each generation can make improvements on the learned information inherited from their parents' generation \citep{D14,K14}.

Genetic evolution and cultural evolution are parallel processes that optimize information in a population.  It is common to consider the interaction between these parallel processes.  Two effects have been well discussed with respect to learning:  the \textit{hiding effect} is when learning shields genetics from selection pressure, thus slowing the evolutionary process, and the Baldwin effect is when learning stimulates genetics, increasing particular selection pressures, and thus speeding up evolutionary adaptation \citep{SSE,PKS,B96}.  These effects help describe the interaction between these parallel processes when they cooperate to improve fitness.  

Another way we can compare genetic and cultural evolution is according to the direction of the evolutionary trajectory.  It is possible for the cultural evolutionary trajectory to diverge from the biological evolutionary trajectory.  In particular, this means that the culture may evolve in directions that are neutral or even detrimental to the biological imperatives of an agent or its genes.  The biological imperatives of genes are merely survival and reproduction \citep{d95,D76}.  A divergent culture may be one that impedes an agents' abilities to survive and/or reproduce.

A simple non-human example of divergent evolutionary trajectories is sexual selection.  Females could select for traits that correlate with fitness.  In this case the culture and the genetic evolution agree.  However, females could select for traits that do not correlate with fitness or correlate negatively with fitness.  In certain birds of paradise the power of female selection has frustrated the fitness of males.  Technically, this is not a case of cultural divergence since in sexual selection divergence occurs between two genetic evolutionary trajectories (male and female) within a single species.  

Some human cultures may frustrate the reproductive or survival capabilities of some of their members (usually for the apparent benefit of the culture).  For instance, a Catholic priest will abstain from reproducing according to the rule of his culture.  Also, a Samurai might kill himself out of shame for failing to meet his cultural obligations.  These cultural practices impede the genetic imperative of the genes in these individuals.  

An extreme and rare case of cultural divergence would be a case where every individual of the culture engages in detrimental cultural practices.  This would include cases of mass abstinence or mass suicide.  Mass abstinence was a cultural belief of the Shakers (in the 1770s-1780s).  Some mass suicides are caused for fear of capture by an enemy culture (as in Masada, Israel in 74 CE or in Pilenai, Lithuania in 1336 CE).  Others are caused by  a cultural belief that the suicide will grant reward in the afterlife (as with the Heaven's Gate mass suicide in 1997) or avoid punishment in this life (as in Jonestown in 1978).

We believe divergent cultural evolution requires at least two properties.  First, genetic and cultural information must be stored in separate information stores.  This rules out models with horizontal transfer of genetic material.  Second, horizontal transfer of cultural information occurs between individuals of the same generation or across generations.  This property rules out evolutionary development models where phenotypic information is not transferred between individuals.  

We believe these are necessary conditions and we believe they are probably not sufficient conditions.  It is difficult to test this hypothesis since our implementation has many other implicit conditions that may play an important role.  Our experiment is designed to test whether these conditions (plus implicit others) can lead to divergent cultural evolution in our population.

In \cite{MC14} we demonstrated a simple instance of divergent genetic and cultural evolution in a population with these properties.  The experiment involved a simple optimization problem, asexual agents, and no spatial environment.  Agents in that experiment showed accelerated optimization and divergence of selection pressures for particular genes and memes.

We have reproduced this experiment in a virtual environment representing real space.  Our agents engage in sexual reproduction and are subject to natural selection.  Our first experiments in this environment involved simple agents with no learning capabilities \citep{MC15a, MC15b}.  We have augmented these agents with individual and social learning mechanisms.


\section{Divergent Cultural Evolution}

In \cite{MC14} we implemented a simple proof of concept and demonstration of divergent cumulative cultural evolution.  Agents in our model engage in all three modes of adaptation: phylogenetic, ontogenetic, and sociogenetic.  Phylogenetic adaptation is adaptation by genetic evolution.  Ontogenetic adaptation is lifetime adaptation.  Sociogenetic adaptation is the result of exchanging learned material between agents.

\begin{figure}[!t]
\begin{center}
\includegraphics[width = 240px]{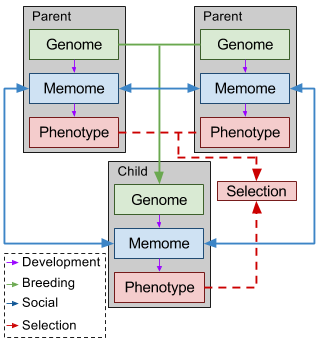}
\caption{The dual inheritance model.}
\label{fig:dim}
\end{center}
\end{figure}

The dual inheritance model (see Fig.~\ref{fig:dim}) describes how these three modes of adaptation interact to create the agent \citep{MC16b}.  Genetic information is inert over the lifetime of the agent in our model.  It is transmitted vertically (from parent to child) during reproductive events and is responsible for the creation of initial cultural (memetic) information that is active during an agent's life.  Memetic information is used during the lifetime of the agent to select a behavior for a particular situation and this information is also adaptive.

As agents are units of selection in our simulation natural selection occurs on the lifetime behavior of an agent (i.e. its phenotype).  This behavior is determined by an interaction of an agent's genome, memome and environment.  As a result both genetic and memetic information is important in determining if an agent lives and reproduces.  In our model memetic selection also occurs.  However, this selection is carried out by agents when they select what information to use, what information to share and whether or not to share their information.

There are two important ways that divergence can occur between genetic and cultural evolution.  It is common that evolutionary trajectories in both the genetic and cultural realm are aligned.  This is common when they are both trying to optimize a behavior.  In these cases it is expected that cultural optimization of the behavior will outpace genetic optimization primarily due to the different the different timescales of these adaptive mechanisms.  The only divergence here is in terms of the rate of optimization.  We call this divergence under \emph{cooperative selection pressures}.

The second type of divergence occurs when genetic selection pressures and cultural selection pressures are contrary.  For instance, sexual reproduction is favored by genetic selection but suppressed in many (human and non-human) cultures.  We call this divergence under \emph{competitive selection pressures}.

We believe that both types of divergence require the properties stated above.  That is, genetic and cultural information must be separate and cultural information must be transmitted horizontally.  We will test our implementation for both types of divergence.

\section {Model}
We have improved on our proof of concept by placing our agents in an environment in which they compete for resources and are subjected to a form of natural selection (i.e. compete for mates) instead of artificial selection (i.e. face a fitness function).  

Our agents live in a random geometric network of resource sites \citep{P03}.  Random geometric networks are an approximation of two dimensional physical space.   At each site agents can spend time gathering the resources available at that site.  Sites in our current model have one, two or three resources available to an agent that gathers at that site.  Agents in our model have genetically or memetically encoded strategies for gathering at a site and the strategy determines the energy cost to the agent.  The energy cost is always at least the number of resources available at that site (one, two or three) and at most five.

Agents have a simple metabolism in which resources are converted into energy.  Energy is used to move around the environment, gather resources, and perform actions like breeding, learning and social learning.  Additional small daily energy penalties are administered for idle activity, old age, and length of genome (only during reproduction).  At the end of each day an agent has a net gain or loss of energy and this contributes to whether the agent lives or dies and whether it has enough energy to reproduce.

\subsection{Genome}
As mentioned above, in the dual inheritance model an agent's genome is inert during its lifetime and therefore is not adaptive nor directly active in behavior selection.  The primary purpose of its genome is to spread genetic information in reproductive events.  The secondary purpose of its genome is to produce an agent's memome upon its birth.

A genome of an agent represents a path of resources sites in the random geometric network.  At each site on this path is also encoded possible behaviors for an agent at that site.  That is, a genome represents a single long path through the network and the actions an agent might take at each site.  

\begin{figure}[!t]
\begin{center}
\includegraphics[width = 240px]{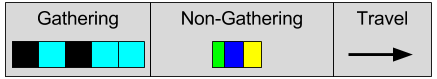}
\caption{A typical gene consisting of gathering, non-gathering, and travel components.}
\label{fig:gene}
\end{center}
\end{figure}

We call each site in this path a \emph{gene} in the genome.  
A typical gene consists of three parts.  A gathering component encodes a strategy for gathering resources at the respective site.  A non-gathering component encodes the energy spent on non-gathering actions like breeding, learning and socializing.  Energy in our model correlates with time (except in reproduction).  In general, more time (energy) spent breeding, learning or socializing increases the likelihood of these activities being successful (more on this below).  Finally, a gene has a travel component which encodes the energy cost of traveling to the next site in the path.

Breeding in our model occurs by sexual reproduction so breeding is a social activity.  In our environment two agents must be at the same site at the same time to breed.  If both agents are performing the breed action for overlapping periods of time then a successful sexual reproduction occurs.  We can see that more time spent breeding during the day will increase an agent's chances of reproducing.  However, spending time breeding comes at an energy cost to the agent as well so it can't afford to spend all of its time breeding.

During sexual reproduction an offspring's genome is created as a recombination of its parents' genomes.  Recombination uses the longest common subsequence of the two genomes.  The offspring's genome also has an opportunity to mutate in this process (see \cite{MC15b} for more details of genetic mechanisms).  Only in these cases is a genome active during an agent's lifetime.

At birth each agent's genome creates a memome.  A memome consists of a collection of memeplexes.  Each memeplex in our model represents a possible set of activities for a single day.  Memeplexes are subsequences of a genome.  During memome generation we start at each gene in a genome and we copy gene by gene into a memeplex.  This continues until the total energy of a segment approaches the maximum daily energy for an agent.  If copying the next gene would exceed the maximum energy the segment is complete and the memeplex is stored.  Memeplexes are stored in the memome along with other memeplexes starting with the same initial site for behavior selection (see below).  
 
Additionally, segments are copied in a backwards direction from every gene.  This means every gene in a genome is responsible for two memeplexes in its memome except the endpoints that are responsible for only a single memeplex.  Notice that since every site in the environment is not necessarily represented in a genome there may be sites that do not have corresponding memeplexes.

\subsection{Memome}

Our agent's cognitive model is inspired by the pandemonium model \citep{J87, F97,MPD10}.  Each memeplex is a sub-path of a genome and thus is a path in the random geometric network.  A memeplex represents a single day's worth of activities.  In the pandemonium model a memeplex is referred to as a daemon.  Daemons compete for control of an agent and in our model memeplexes compete for control of an agent.

Behavior selection is also quite similar to the MAP-elites strategy for multi-objective evolutionary optimization \citep{mc15}.  We have memeplexes organized in the memome based on the starting site.  An agent's day begins by selecting all memeplexes in its memome that begin at the agent's current site.  Recall that these memeplexes represent a full day's worth of activities.  The memeplex from this set that rewards the maximum (expected) resources for the day while minimizing the energy expenditure is selected.  

This is the primary force of cultural selection in our current implementation.  It means that memeplexes with maximum resource-to-energy ratio are selected as behaviors.  Since in our social learning mechanism agents that engage in social learning share only the current day's memeplex it means this selection mechanism is also used to select which memeplexes are shared during social learning.

An agent can engage in individual learning only if its selected memeplex includes at least one meme that has a non-zero learning component.  This means that an agent must spend time engaged in learning at at least one site during a day.  When this occurs an agent will clone its memeplex for the day and apply a mutation.  The new memeplex is added to its memome.  This allows an agent to possibly generate a memeplex that is more efficient at the same activities or generate an alternative sequence of activities.  

We can see that having non-zero learning components in memes would benefit the agent.  However, spending time learning during the day comes at an energy cost, as with breeding.  Further, as our current implementation only allows a single learning event in a day it is not beneficial for an agent to spend more than the minimum amount of time learning.

An agent can engage in social learning once per day.  The process for social learning is very similar to the process for sexual reproduction.  Social learning can only occur if an agent spends time engaged in a social learning action at least one site during a day.  However, for social learning to occur, another agent must also be at the same site at the same time engaged in social learning.  If this occurs then the two agents will swap mutated copies of the memeplexes they used for that day.  We treat this mechanism as roughly equivalent to telling the other agent what they did for the day.

Again we see a benefit to having social learning in memes as this will increase the chances of an exchange of memeplexes.  However, as with learning and breeding, an agent cannot afford to spend too much time performing social learning during a day.

Both of these learning mechanisms allow for new memeplexes to be added to the memome which means an agent can adapt its behavior.  When it begins its day at the same site again it may select one of the new memeplexes instead.

\section{Experimental Setup}

We conduct three similar experimental runs with agents of different capabilities.  The first control group we call breeders, and while they still have memomes, their learning and social learning mechanisms have been turned off.  In these agents, learning and social learning components of genes/memes are still present but inactive.  The second control group we call learners.  They are similar to breeders as they still lack social learning mechanisms, but they have their individual learning mechanisms intact.  Like the breeders, they still have social learning components of genes/memes but they are inactive.  The third group is our experimental group.  We call them socializers and they have all the functionality described above.

Each run is seeded with one hundred randomly generated agents.  All genes in the randomly generated genome have learning and social learning components initialized to zero.  Since it would be impossible for agents to breed if this were true of breeding components, we instead have a chance of initializing breeding components to non-zero values.  As a result, agents must mutate learning and social learning in order to take advantage of these mechanisms.  We allow our simulations to run for 5000 days.  Under these settings every initial population is viable although some of our socializer simulations terminate early due to a catastrophic colony collapse (see below).

We gather data on many aspects of our agents' lives.  In particular we gather information on the proportion of a genome or memeplex devoted to breeding, learning or social learning.  Recall that genomes and memeplexes are both paths of sites and they can vary in length from agent to agent.  We can measure the length of a genome or memeplex in two ways.  We can count the number of genes/memes (representing sites) or we can measure the energy cost of a genome/memeplex as a whole.  We record the proportion of a genome/memeplex devoted to breeding as the total energy devoted to breeding in a genome/memeplex over the total energy cost of the genome/memeplex.  We do the same for learning and social learning.

We use a slightly different method to calculate the optimization of a genome/memeplex.  For each site there is a gathering action.  Recall the action will cost energy at least the number of resources rewarded at a site (one, two or three) and at most five.  We can count all energy used above the minimum as wasted energy.  For each genome/memeplex we compute the average wasted energy per gene/meme.

The control groups of breeders and learners should not display divergent cumulative cultural evolution.  Among the control groups we expect differences in  genome and memeplex measurements but we expect these differences to be small.  In the socializers we  expect to see evidence of divergent cumulative cultural evolution.  We expect memeplexes will show evidence of greater optimization than the genomes, at least after enough time for cumulative cultural evolution to emerge.  Recall this is a case of divergence under cooperative selection pressures.

We also expect to see divergence under competitive selection pressures (i.e. when they pull in different directions).  We expect that genetic selection will select for breeding while also having indirect selection for learning and social learning.  We expect that there will remain some selection pressure for social learning and learning in the memeplexes, but the pressure to optimize these actions out of a memeplex will also be strong.  Breeding is not to the advantage of the memetic selection mechanism so we expect it to be selected against by memetic selection.  

\section{Observations and Discussion}

Our experiment was replicated 130 times on a variety of random geometric networks.  All data presented in this section is averaged over these 130 runs.


We wish to begin with a discussion of an apparent slowdown of genetic evolution caused by cultural evolution.  When we first investigated the breeding, learning and social learning genes in the genomes of socializers we found that gene concentrations for these components grow at the same rate for our two control groups but at a slower rate for the socializers.  We thought this could be due to a  hiding effect occurring between cultural evolution and genetic evolution.

\begin{figure}[!t]
\centering
\includegraphics[width = 240pt]{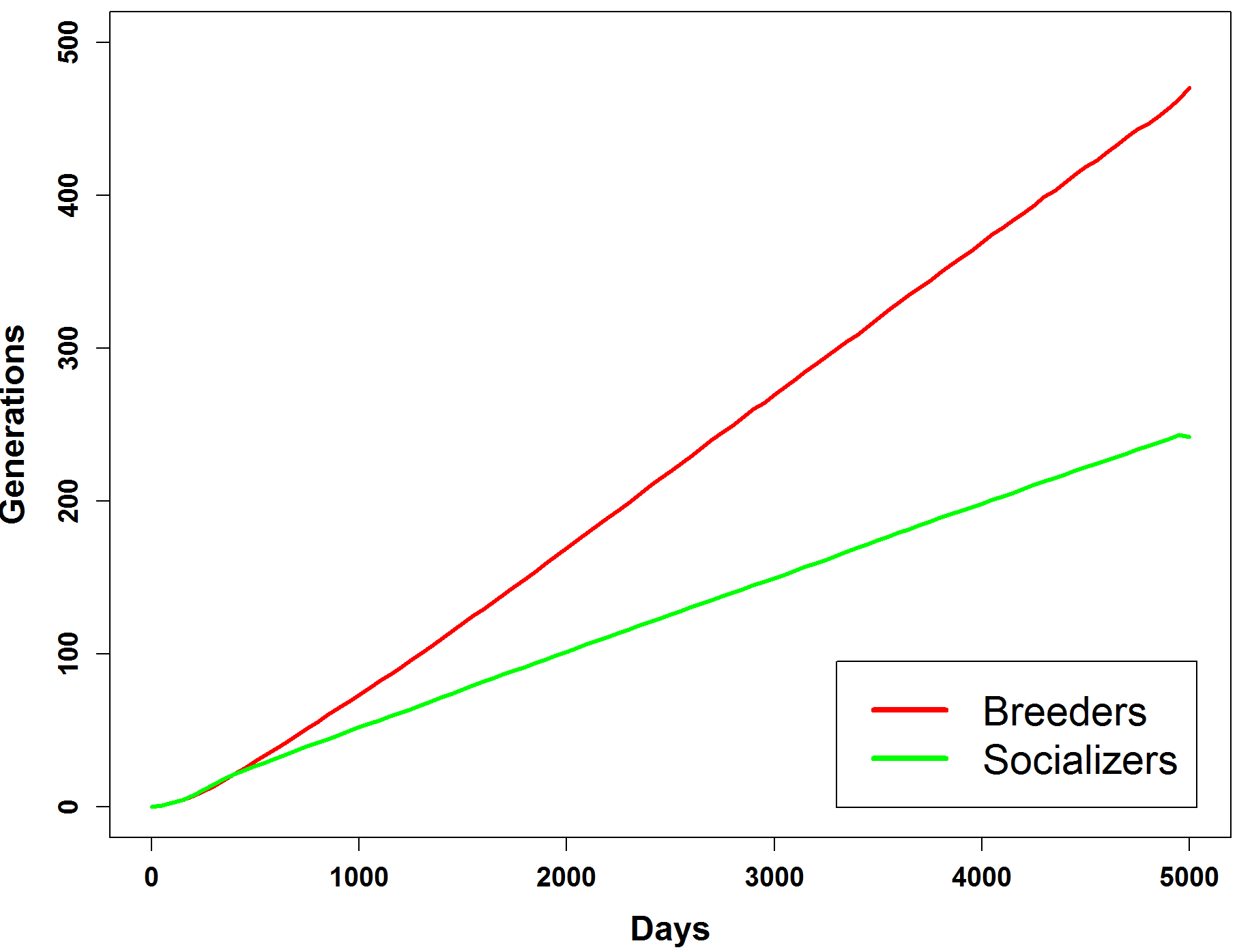}
\caption{Generation over time for breeders and socializers.  Generation is defined as the maximum generation in the population.  A child's generation is one greater than the max of its parents' generations.}
\label{fig:gens}
\end{figure}

We saw a similar effect on genome length over time and other data we gathered.  However with further investigation we were able to determine the source of the slowdown.  In both control groups the generation of agents increased at identical rates.  The generations of socializers increased at about half the rate (Figure \ref{fig:gens}).  This is due to an emergence of eusocial breeding culture in our agents \citep{MC16b}.

We think this evidence suggests that cultural evolution can slow genetic evolution over time, but possibly not over generations.  That is, the shielding of genetic selection pressure is not really there.  Instead there is a selection pressure for longer generations which has the result of slowing genetic evolution over time.

\begin{figure}[!t]
\centering
\includegraphics[width = 240pt]{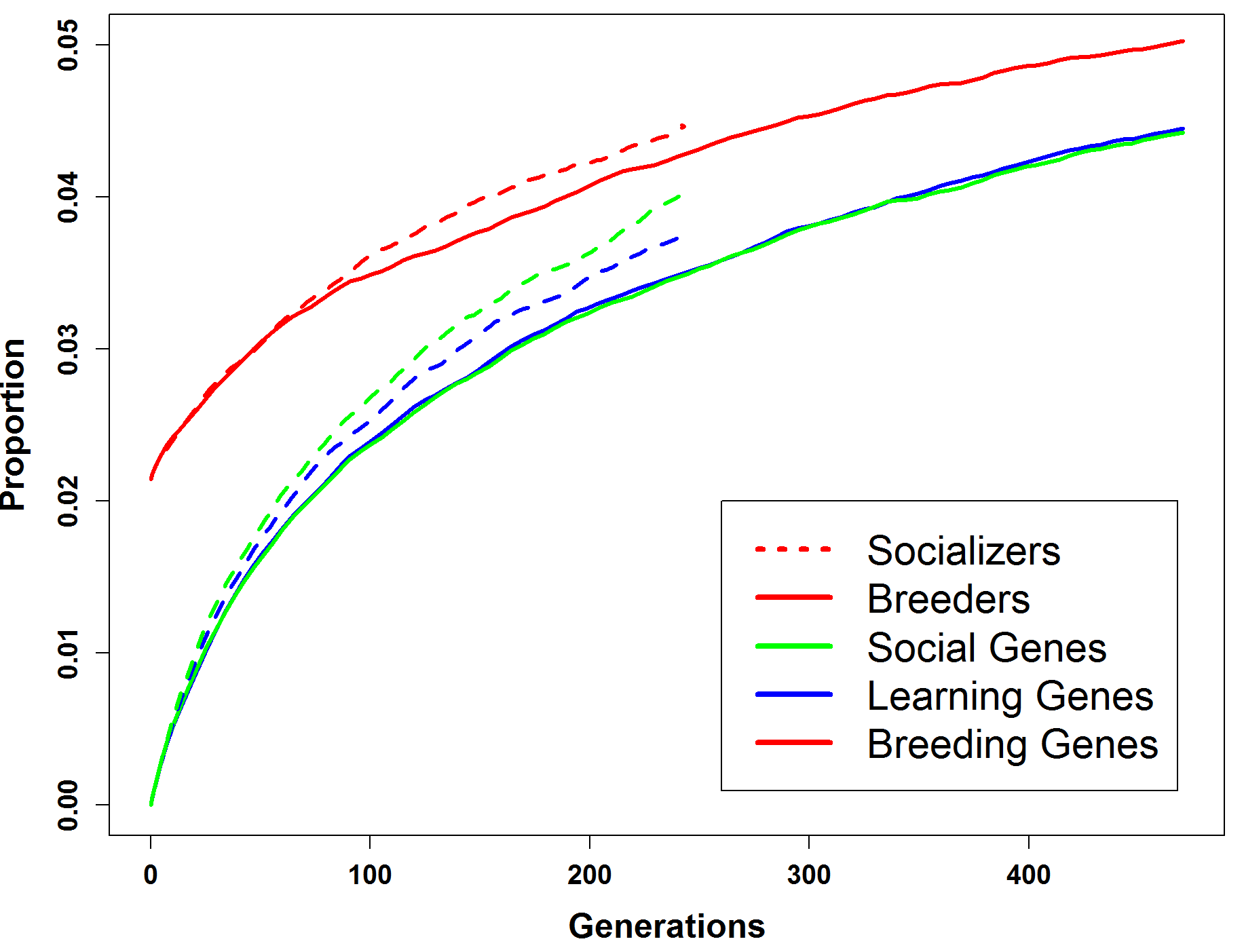}
\caption{Concentration of breeding, learning, and socializing genes in the genomes of breeders (dotted) and socializers (solid) over generations.}
\label{fig:genes}
\end{figure}

To confirm this we plotted gene concentrations over generations instead of days (Figure \ref{fig:genes}).  We can see that the socializers actually have weak acceleration of evolution over time according to this plot.  These differences are small.  We suspect that a greater significant difference might occur if we allowed our simulation to run for more days.

\begin{figure}[!t]
\centering
\includegraphics[width = 240pt]{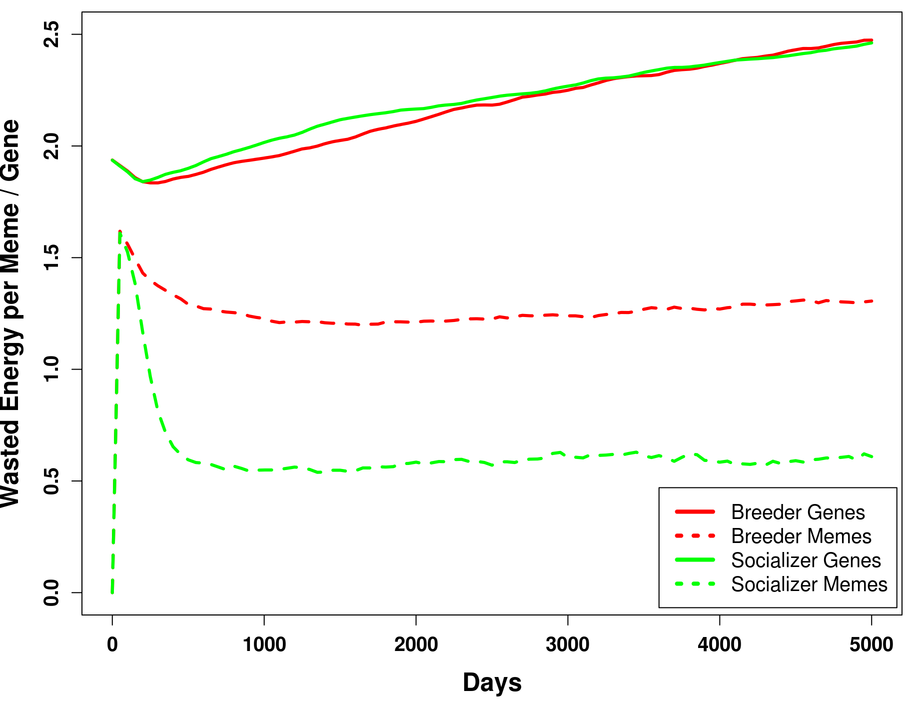}
\caption{Average wasted energy per gene/meme in the genomes and memomes of breeders and socializers.}
\label{fig:opt}
\end{figure}

Now we wish to consider evidence for divergence under cooperative selection pressures.  So we turn  to a discussion of the relative optimization of genomes and memeplexes.  Breeders and socializers showed a slight trend to less optimized genomes over time (Figure \ref{fig:opt}).  This is not unexpected as much of the genome is not actually used during the agent's lifetime and thus is not subjected to selection at all.  Further the genomes of agents increase overtime so this increases the size of the unused region.  This has a result of stagnation of optimization in the genome except for a very small region.  We can see in the control groups that this region is indeed optimized.  We see this in the measurement of optimization of the daily memeplexes.  The memeplexes selected for activity are more optimized than the genome as a whole.  The memeplexes in the control runs undergo an early stage of optimization before stagnating.

In the socializers there is also an early stage of optimization before stagnation.  However the stage of optimization is considerably greater in socializers than non-socializers.  Stagnation in the control runs is in part due to a weak genetic selection pressure for optimization.  Selection pressure for optimization in the memome is stronger and most evolved memeplexes are highly optimized (nearing zero wasted energy).  However the data shows an average of all agents in the population.  Only older agents have the evolved memeplexes and the younger agents have memeplexes that are still just close copies of regions of their genomes.  When we average over all agents we will never optimize to zero.

\begin{figure}[!t]
\centering
\includegraphics[width = 240pt]{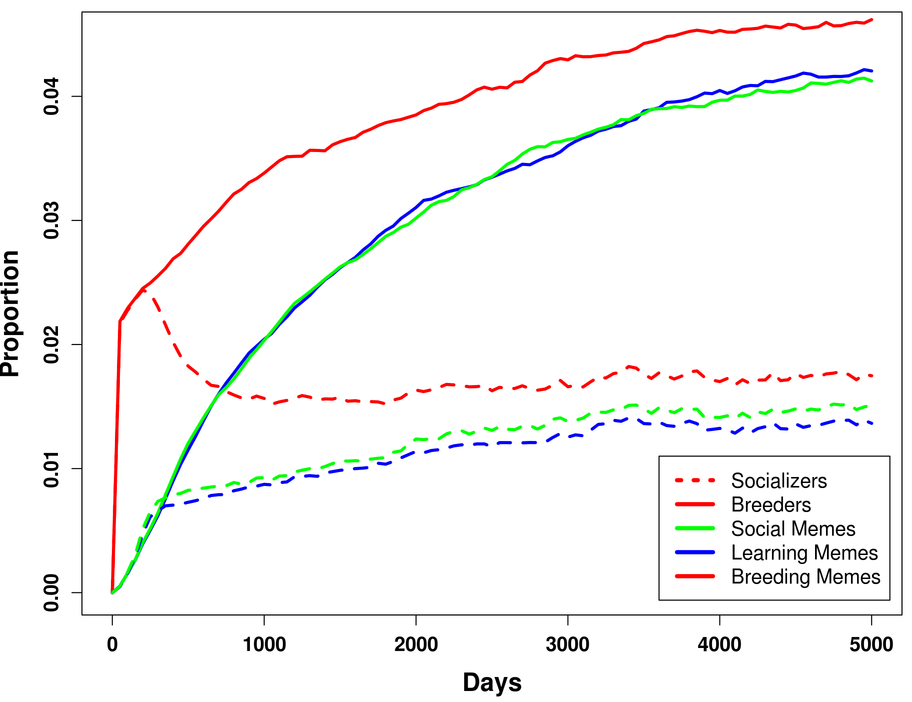}
\caption{Concentration of breeding, learning, and socializing memes in the memomes of breeders (dotted) and socializers (solid).}
\label{fig:memes}
\end{figure}

Now let's consider divergence under competitive selection (see Fig~\ref{fig:memes}).  The strongest competitive pressure is relative to breeding actions.  Treating memeplexes as a type of parasitic organism we can see that they are only concerned with their replication into new hosts.  They are not concerned with their host's reproduction, even if this leads to fewer available hosts in the long run.  A common end for this type of parasite is to die off after killing all available hosts. 

We observe that soon after social learning emerges in the population memeplexes diverge from the genome.  Breeding actions in a genome continue to be selected for, but breeding actions in a memome are selected against and their concentration decreases before stagnating.  Notice again this stagnation is due in part to the averaging over the population.  Some agents still are young and have a higher concentration of breeding actions than more evolved memeplexes.

We observe that in many socializers, breeding actions in their memeplexes are reduced to zero.  This is also clear by noting that about 54\% of breeders and learners have children while only 36\% of socializers do.  Memeplexes co-opt the agent for their reproductive ends instead of the genome's reproductive ends.  

Interesting cases of collapsed colonies occur when these memeplexes spread to all agents before they can breed.  In many runs of socializers we witness isolated colonies completely dying out.  Out of 100 runs 21 ended when all agents died out before the 5000 day limit is reached.  This only occurs in socializers.

In breeders and learners, colonies can face extinction due to a shortage of resources caused by overpopulation.  In these situations agents can't get enough resources to reproduce and in some cases can't get enough to survive.  However this situation cures itself as agents die out.  As agents die, they no longer collect resources.  The resources can instead go to the young and eventually young agents can reproduce.  This causes cycles in population density but never a collapse.

Consider learning and social learning actions.  There is strong early selection pressure in genomes and memomes for these actions.  However as social learning kicks in, memetic selection against wasting time on these action takes over.  Remember it is beneficial to spend as little time as possible on learning and so optimized memeplexes spend energy learning at only a single site.  Many optimized memeplexes do not spend energy on learning at all.  This is a detriment but if a memeplex is already highly optimized there is little benefit to learning.  Further if a memeplex still spends time social learning an agent can still be adaptive.

It is also beneficial to not waste time on social learning.  As above when social learning begins optimization places a negative selection pressure on social learning actions.  However it is a memeplex's responsibility to spread itself.  A memeplex that evolves to spend no time on social learning will never be spread (but may still be useful to an agent).  Thus, we usually observe at least one site in a memeplex with social learning time and often more.  This means that a memeplex usually has a particular site at which it spreads itself to other agents.  This builds cultures of agents around meme spreading sites. 

We notice for all breeding, learning and social learning actions as well as for optimization, cumulative cultural evolution causes a divergent effect from genetic evolution.  In our simulations, this period of time occurred early on (in the first 500 days) and then social learning maintains an optimized culture through spreading evolved memeplexes to new agents.

\begin{figure}[!t]
\centering
\includegraphics[width = 240pt]{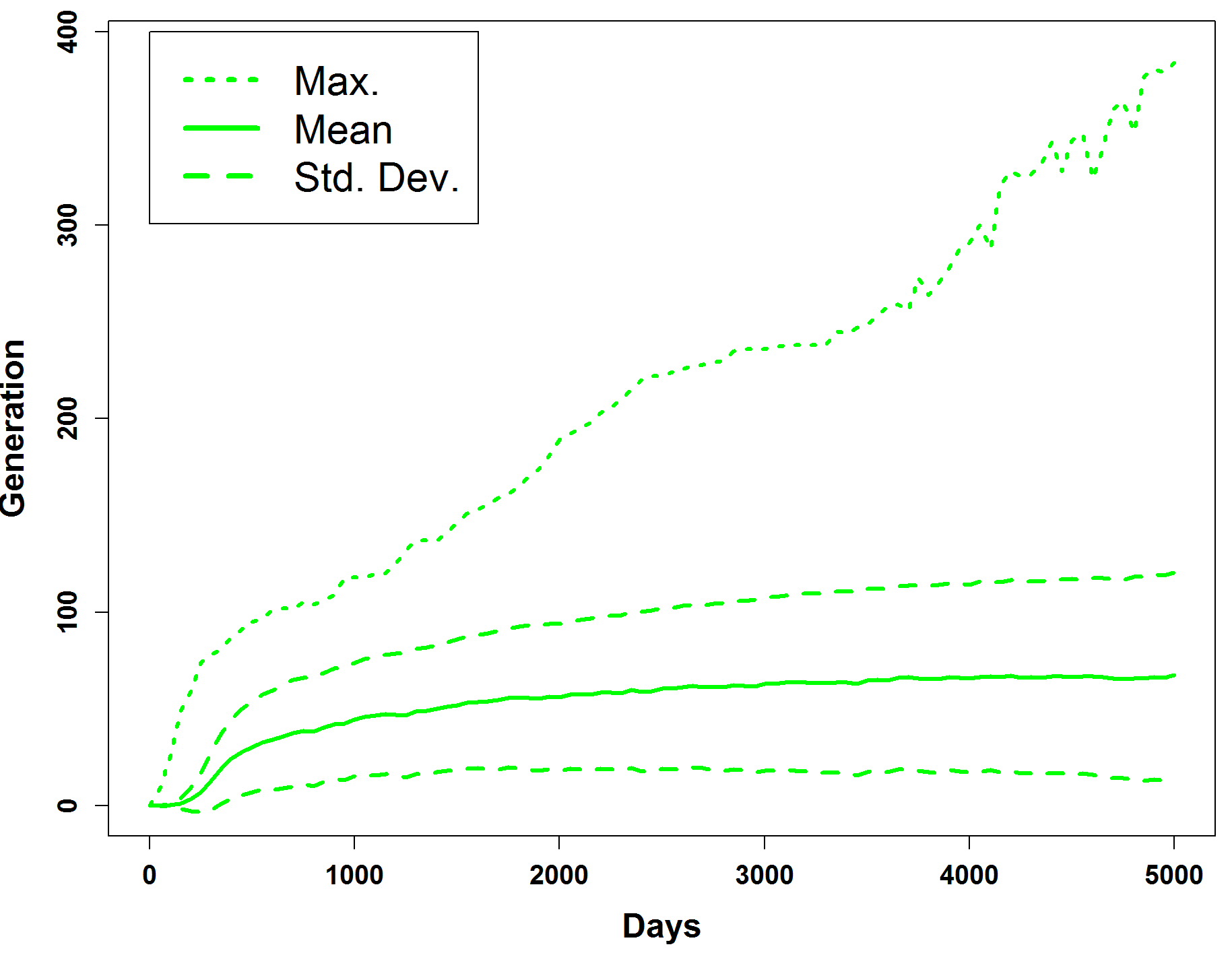}
\caption{Average and maximum generation of active memeplexes over time with one standard deviation around the mean.}
\label{fig:memegens}
\end{figure}

To track cumulative cultural evolution we also assigned a generation to each memeplex.  All original memeplexes created at birth are assigned generation zero.  Whenever it is cloned in learning or social learning the clone is assigned generation one higher than its parent.

We can detect cumulative cultural evolution by detecting an increase in memeplex generation over time, especially from generation to generation (Figure \ref{fig:memegens}).  We can indeed confirm cumulative cultural evolution in our socializers by this method.  Of course breeders can only have a memeplex generation of zero.  Learners however can have a memeplex generation above zero if they learn better memeplexes.  However these memeplexes can never leave their initial host and die with the host.  Therefore, no cumulative culture can accrue \citep{K14}.  We do see that learning agents can maintain a low non-zero memeplex generation but not one that increases over time.

Interestingly, in the socializers we do notice two stages in memeplex generation growth.  The initial stage is more rapid and the second stage grows slowly at a fixed rate over time.  We note that the period of rapid growth coincides with the period in which the memeplexes are being optimized prior to stagnation.  The slower growth rate coincides with the stage where agents pass around optimized memeplexes. 

\section{Conclusions}

Our implementation incorporated two information stores, one for genetic information and one for cultural information.  It also had mechanisms of horizontal transfer of cultural information between agents of multiple generations.  Our implementation is an example of our dual inheritance model of cultural evolution \citep{MC16a}.  Our implementation demonstrates divergent cumulative cultural evolution under both conditions of cooperative and competitive selection pressures.

We can see that populations of agents that participate in the dual inheritance model can accelerate optimization relative to selection pressures that are cooperative between the genetic and memetic world.  This is due to the potential for many cultural generations in a single biological generation.  Thus, optimization can occur much faster over real time in cultural evolution than in biological evolution.  Further, since the selection pressure on both the genome and the memome operate in the same direction there cannot be conflict between these pressures.  The only divergence in this case is in terms of speed to convergence.

We also see cases where the selection pressure on the genome operates in an opposite direction for the memome.  Genes and memes care only for spreading themselves.  For genes, spreading occurs through reproductive events but for memes spreading occurs through social learning events.  So it is not surprising that genes that increase the success of reproductive events are selected for by biological evolution and memes that increase the number and success of social learning events are selected for by cultural selection.

The contrary is not necessarily true.  Genes that increase the number and success of social learning events may be selected for by biological evolution if social learning also helps improve survival or reproductive success.  This is true in our simulation.  Memes that increase the number and success of reproductive events have only a distant and indirect impact on the number and success of social learning actions.  Since they also have a detrimental effect on the optimization of the memeplex there is a considerably stronger selection to avoid these actions.

Finally, we still see an interesting divergence in behavior of young (inexperienced) agents and old (experienced) agents.  Young agents have had very little or no time to adapt their initial set of memeplexes either through individual optimization or through learning from others.  Thus their behavior is still largely determined by their genome, which may also be the case in humans \citep{T16,CG11}.  This means they are more likely to breed, learn and social learn than old agents because all of these actions occur in much higher concentrations in the genome than in the memome.

The interesting impact of these trends is that young agents are more likely to be parents (i.e. before they learn better).  They are more likely to learn from the environment more than older experienced agents.  Finally, they are also more social.  They are more likely to seek out social learning events than old agents.  We find this conclusion interesting for two reasons.  First, we see an emergent organization in our populations around age.  Second this organization mimics the same organizations in other models and natural populations \citep{L13,TM09}.  Young humans are more likely to have children, more likely to attempt to improve themselves through learning, and more likely to seek out the knowledge of others than their older counterparts \citep{D12,H11}.  Further older humans that engage in social learning are more often teachers than learners and this is also born out by our experiment.   

Finally we believe that the observed divergent cumulative cultural evolution is due to a critical component of the dual inheritance model.  Specifically we think it is critical to keep genetic and cultural information separate from one another even if they store the same kinds of information (as in our implementation).  Without separate information stores there is no environment for divergence to occur.  Secondly it is important that cultural information can be transmitted between members of the same generation and between generations.

\section{Acknowledgments}

The authors would like to thank the advice of anonymous reviewers.  Jobran Chebib was supported by the Swiss National Science Foundation (grant PP00P3\_144846/1 awarded to Fr\'ed\'eric Guillaume).

\footnotesize
\bibliographystyle{apalike}
\bibliography{example}

\begin{thebibliography}{}

\bibitem[Baldwin, 1896]{B96}
Baldwin, J.~M. (1896).
\newblock A new factor in evolution.
\newblock {\em The american naturalist}, 30(354):441--451.

\bibitem[Boyd and Richerson, 1996]{br96}
Boyd, R. and Richerson, P.~J. (1996).
\newblock Why culture is common, but cultural evolution is rare.
\newblock In {\em Proceedings-British Academy}, volume~88, pages 77--94. OXFORD
  UNIVERSITY PRESS INC.

\bibitem[Csibra and Gergely, 2011]{CG11}
Csibra, G. and Gergely, G. (2011).
\newblock Natural pedagogy as evolutionary adaptation.
\newblock {\em Philosophical Transactions of the Royal Society of London B:
  Biological Sciences}, 366(1567):1149--1157.

\bibitem[Dawkins, 1976]{D76}
Dawkins, R. (1976).
\newblock {\em The selfish gene}.
\newblock Oxford University Press.

\bibitem[Dean et~al., 2014]{D14}
Dean, L.~G., Vale, G.~L., Laland, K.~N., Flynn, E., and Kendal, R.~L. (2014).
\newblock Human cumulative culture: a comparative perspective.
\newblock {\em Biological Reviews}, 89(2):284--301.

\bibitem[Demps et~al., 2012]{D12}
Demps, K., Zorondo-Rodr{\'\i}guez, F., Garc{\'\i}a, C., and Reyes-Garc{\'\i}a,
  V. (2012).
\newblock Social learning across the life cycle: cultural knowledge acquisition
  for honey collection among the jenu kuruba, india.
\newblock {\em Evolution and Human Behavior}, 33(5):460--470.

\bibitem[Dennett, 1995]{d95}
Dennett, D.~C. (1995).
\newblock Darwin's dangerous idea.
\newblock {\em The Sciences}, 35(3):34--40.

\bibitem[Franklin, 1997]{F97}
Franklin, S. (1997).
\newblock {\em Artificial minds}.
\newblock MIT press.

\bibitem[Henrich and McElreath, 2003]{hm03}
Henrich, J. and McElreath, R. (2003).
\newblock The evolution of cultural evolution.
\newblock {\em Evolutionary Anthropology: Issues, News, and Reviews},
  12(3):123--135.

\bibitem[Hewlett et~al., 2011]{H11}
Hewlett, B.~S., Fouts, H.~N., Boyette, A.~H., and Hewlett, B.~L. (2011).
\newblock Social learning among congo basin hunter--gatherers.
\newblock {\em Philosophical Transactions of the Royal Society of London B:
  Biological Sciences}, 366(1567):1168--1178.

\bibitem[Jackson, 1987]{J87}
Jackson, J.~V. (1987).
\newblock Idea for a mind.
\newblock {\em ACM SIGART Bulletin}, (101):23--26.

\bibitem[Kempe et~al., 2014]{K14}
Kempe, M., Lycett, S.~J., and Mesoudi, A. (2014).
\newblock From cultural traditions to cumulative culture: parameterizing the
  differences between human and nonhuman culture.
\newblock {\em Journal of theoretical biology}, 359:29--36.

\bibitem[Lehmann et~al., 2013]{L13}
Lehmann, L., Wakano, J.~Y., and Aoki, K. (2013).
\newblock On optimal learning schedules and the marginal value of cumulative
  cultural evolution.
\newblock {\em Evolution}, 67(5):1435--1445.

\bibitem[Lindblom and Ziemke, 2003]{LZ}
Lindblom, J. and Ziemke, T. (2003).
\newblock Social situatedness of natural and artificial intelligence: Vygotsky
  and beyond.
\newblock {\em Adaptive Behavior}, 11:79--96.

\bibitem[Marriott and Chebib, 2014]{MC14}
Marriott, C. and Chebib, J. (2014).
\newblock The effect of social learning on individual learning and evolution.
\newblock In {\em The Fourteenth Conference on the Synthesis and Simulation of
  Living Systems}, pages 736--743. MIT Press.

\bibitem[Marriott and Chebib, 2015a]{MC15a}
Marriott, C. and Chebib, J. (2015a).
\newblock Emergence-focused design in complex system simulation.
\newblock In {\em European Conference on Artificial Life}. MIT Press.

\bibitem[Marriott and Chebib, 2015b]{MC15b}
Marriott, C. and Chebib, J. (2015b).
\newblock Finding a mate with no social skills.
\newblock In {\em Proceedings of the 2015 conference on Genetic and
  Evolutionary Computation}. ACM.

\bibitem[Marriott and Chebib, 2016a]{MC16a}
Marriott, C. and Chebib, J. (2016a).
\newblock Finding a mate with eusocial skills.
\newblock In {\em Proceedings of the 2016 Conference on the Synthesis and
  Simulation of Living Systems}. MIT Press.

\bibitem[Marriott and Chebib, 2016b]{MC16b}
Marriott, C. and Chebib, J. (2016b).
\newblock Modelling the evolution of gene-culture divergence.
\newblock In {\em Proceedings of the 2016 Conference on the Synthesis and
  Simulation of Living Systems}. MIT Press.

\bibitem[Marriott et~al., 2010]{MPD10}
Marriott, C., Parker, J., and Denzinger, J. (2010).
\newblock Imitation as a mechanism of cultural transmission.
\newblock {\em Artificial Life}, 16:21--37.

\bibitem[Mesoudi et~al., 2006]{MWL}
Mesoudi, A., Whiten, A., and Laland, K. (2006).
\newblock Towards a unified science of cultural evolution.
\newblock {\em Behavioral and Brain Sciences}, 29:329--383.

\bibitem[Mouret and Clune, 2015]{mc15}
Mouret, J.-B. and Clune, J. (2015).
\newblock Illuminating search spaces by mapping elites.
\newblock {\em arXiv preprint arXiv:1504.04909}.

\bibitem[Paenke et~al., 2006]{PKS}
Paenke, I., Kawecki, T., and Sendhoff, B. (2006).
\newblock On the influence of lifetime learning on selection pressure.
\newblock In Rocha, L., Yaeger, L., Bedau, M., Floreano, D., Goldstone, R., and
  Vespignani, A., editors, {\em Artificial life X}. MIT Press.

\bibitem[Penrose, 2003]{P03}
Penrose, M. (2003).
\newblock {\em Random geometric graphs}, volume~5.
\newblock Oxford University Press Oxford.

\bibitem[Sznajder et~al., 2012]{SSE}
Sznajder, B., Sabelis, M.~W., and Egas, M. (2012).
\newblock How adaptive learning affects evolution: Reviewing theory on the
  baldwin effect.
\newblock {\em Evolutionary Biology}, 39:301--310.

\bibitem[Thornton and Malapert, 2009]{TM09}
Thornton, A. and Malapert, A. (2009).
\newblock Experimental evidence for social transmission of food acquisition
  techniques in wild meerkats.
\newblock {\em Animal Behaviour}, 78(2):255--264.

\bibitem[Tomasello, 2016]{T16}
Tomasello, M. (2016).
\newblock The ontogeny of cultural learning.
\newblock {\em Current Opinion in Psychology}, 8:1--4.

\bibitem[Whiten et~al., 2011]{W11}
Whiten, A., Hinde, R.~A., Laland, K.~N., and Stringer, C.~B. (2011).
\newblock Culture evolves.
\newblock {\em Philosophical Transactions of the Royal Society of London B:
  Biological Sciences}, 366(1567):938--948.

\end{thebibliography}

\end{document}